\documentclass[11pt]{amsart}

\usepackage{amsmath}
\usepackage{amssymb}
\usepackage{amsthm}
\usepackage{amscd}
\usepackage{amsfonts}
\usepackage{graphicx,caption, subcaption}%
\usepackage{fancyhdr, geometry}
\usepackage{url, hyperref, color, verbatim}
\usepackage[normalem]{ulem}

\newtheorem{theorem}{Theorem}[section]

\newtheorem{problem}[theorem]{Problem}

\theoremstyle{definition}
\newtheorem{definition}[theorem]{Definition}
\newtheorem{example}[theorem]{Example}

\theoremstyle{remark}
\newtheorem{remark}[theorem]{Remark}

\numberwithin{equation}{section}

\newcommand{\conv}{conv}

\begin{document}

\title{A survey of discrete methods in (algebraic) statistics for networks} 

\author{Sonja Petrovi\'c}
\address{Department of Applied Mathematics, Illinois Institute of Technology, Chicago, Illinois 60616}
\email{sonja.petrovic@iit.edu}


\date{August 2015; revised January 2016.}



\begin{abstract}
Sampling algorithms, hypergraph degree sequences, and polytopes play a crucial role in statistical analysis of network data. 
This article offers a brief overview of open problems in this area of discrete mathematics from the point of view of a particular family of statistical models for networks called exponential random graph models. 
The problems and underlying constructions are also  related to well-known concepts in   commutative algebra and  graph-theoretic concepts in computer science. 
We outline a few lines of recent work that highlight the natural connection between these fields and unify them into some open problems. While these problems are often relevant in discrete mathematics in their own right, the emphasis here is on statistical relevance with the hope  that these lines of research do not remain disjoint.  
 Suggested specific open problems and general research questions should advance   algebraic statistics theory as well as applied statistical tools for rigorous statistical analysis of networks. 
\end{abstract}

\maketitle

\section{Introduction}

 The development of a rich literature on networks in the past decade has left 
 ample opportunities for complementary 
 mathematically rigorous results that should serve as the foundation for statistical modeling and fast computation of network features.   
In this review, we focus on exponential families for random graphs, models over equivalence classes of graphs summarized by a selected set of graph or network summary statistics. Here, the word `model' is used in the statistical sense, and the terms `graph' and `network' are used interchangeably. 
Introductory paragraphs of Sections~\ref{sec:exactTesting} and \ref{sec:polytopes} offer an answer to the following question: Why consider statistics when studying random graphs and networks? A summarized answer can be phrased as follows: \emph{how} a network is generated is crucial to properly calculate statistical network properties and specify the distributions being sampled. This includes properties that depend on both degrees of nodes in the network, higher-order degree correlations, or other, possibly global, summary statistics; cf.\  an excellent example and the concluding paragraph in  \cite[Section 4.2.1]{JacksonNtwksBook}. 

The reader may wonder why the focus on exponential families for random graphs, or ERGMs.  
While being able to write a random graph model as an exponential family is not  impressive in and of itself,  understanding the geometry, algebra, and the discrete structures supporting the model offer various statistical insights that are the theme of this chapter. 
A recent computer science article \cite{ERGMsAreHard2014} shows that ERGMs are `hard', that is, their normalizing constants are incomputable in general. While this important result about inapproximability in polynomial time formalizes  the inherent complexity of this family of models, as a statistical model family they are still very broad and rich and have desirable  properties. 
As such, they in the very least offer the theoretical foundation for studying random graphs and networks, and a bedrock for algorithmic exploratory analysis of sampling distributions of various graph summary statistics.

The statistics literature on understanding and developing network models is ever-growing; a partial list of references is offered below in context. Within this realm, though, the basic problem of establishing rigorous procedures for statistical inference is still a challenge, due to the varying complexity of the models and types of properties of data they capture, as well as networks being a novel data type in terms of traditional statistics. 

\medskip
The motivating problem for this discussion is thus a basic one of \emph{statistical inference}, tightly related to several fundamental   tasks required for statistical analysis of  (any kind of) data: parameter estimation\footnote{Sometimes called `fitting' in computer science or statistical physics literature, but it is different from model fit testing.}, sampling from the distributions in the model, testing model/data fit, and model selection. 
The broad goal of statistical inference is to decide,  with a high degree of confidence, whether  an observed data  sample $x=x_1,\dots,x_N$ can be regarded as a draw from a distribution $p_{\theta_0}\in\mathcal M$ coming from a candidate statistical model $\mathcal M$, specified by the unknown parameter $\theta_0$.  
Of course, this entails two crucial steps: 1) {estimation}: use the observed data $x$ to produce an optimal estimate $\hat\theta=\hat\theta(x)$ of  $\theta_0$;
and 2) {goodness-of-fit testing}: assess whether $p_{\hat\theta}$ can be considered a satisfactory generative model for the observed data $x$. 
Surprisingly, these fundamental tasks pose a family of open problems, in particular for discrete  sparse small-sample data such as networks \cite{KrivitskyKolaczyk15,hunter2008goodness, HL81, Habe:1981, Agresti, p1asymptotics, NtwkApproxFittingDynamics15, CDS11}. 
These problems and their natural connection to discrete structures are  the focus of this short overview.

In the remainder of the chapter, we will use the $\beta$ model for random graph, defined in Example~\ref{ex:betaModel}, as a running example to illustrate the structure and implications of various statistical  modeling questions that discrete mathematics tools can answer.

\section{Preliminaries} 

We begin with some technical preliminaries on linear ERGMs and statistical considerations about these models. 
A \emph{statistical model} $\mathcal M$ is a family of probability distributions indexed by a set of parameters $\Theta\subset\mathbb R^n$. 
In exponential random graph models, or ERGMs for short, 
 one first selects the network characteristic $T$ of interest in a particular problem. This selection is done so that $T$ represents some interpretable and meaningful summary statistic of a network. The resulting model is then the collection of probability measures
\[
	\mathcal M= \{p_\theta: \theta\in\Theta\},
\] 
indexed by points in $\Theta\subseteq\mathbb R^n$ such that, for any $\theta\in\Theta$, the probability of observing a
given network\footnote{As is standard in statistics, $G$ represents a random variable and the lowercase $g$ its realization.} $G=g$
takes on the exponential form 
\begin{equation}\label{eq:expfam}
	p_\theta(g)=\exp\{\langle T(g),\theta\rangle - \psi(\theta)\}, 
\end{equation} 
where $\psi(\theta):= \sum_{g} \exp\{\langle T(g), \theta\rangle\}$ is a normalizing function called the log-partition function, and $T(g)$ is the vector of minimal sufficient statistics for the model. 
The exponential form of the probabilities above  is a central theme in statistical theory, and statistical models of this form, known as exponential families, are known to exhibit optimal statistical performance and appear with a growing appeal in the machine learning community as well. Consistent with the running examples, we are considering \emph{linear} exponential families for now, that is, $\langle T(g),\theta\rangle$ is a linear map of the state space. General exponential families can use non-linear maps, of course. 
For a more detailed statistical treatment of this family of models the reader is referred to classical references \cite{Brown86}, \cite{Barndorff-Nielsen} and \cite{LehmannCasella83, LehmannCasella98}, as well as a more recent book \cite{BickelDoksum06}. 
ERGMs are, effectively, models over
\emph{equivalence classes of networks}, where two networks $g_1$ and $g_2$ are regarded as probabilistically equivalent whenever $T(g_1) = T(g_2)$. Conditioning on the values of $T$ permits the reduction of the data through sufficient statistics and thus eliminates nuisance parameters \cite{Agresti}, which can be very helpful  in applications.

\begin{example}[The $\beta$ model for graphs]\label{ex:betaModel}
	The $\beta$ model for graphs is a well-known statistical model for  random undirected graphs. 
In its original form, the $\beta$ model considers simple graphs, that is, does not allow loops or multiple edges; but the general  version allows edges to appear with bounded multiplicity. 
It  essentially assigns parameters $\beta_1,\dots,\beta_n$ to the $n$ vertices in the network that measure their `friendliness', or propensity to attract edges. 
	The edges  are  then assumed to be independent and appear with probability proportional to the product of the parameters of its vertices; in symbols: 
	\[
		Prob(G=g) 
		 \propto \prod_{\{i,j\}\in E(g)}\beta_i\beta_j = \prod_{i=1}^n\beta_i^{\deg(i)},
	\]
where `$\propto$' refers to the fact that the resulting quantity should be normalized by the log-partition function. 
In this model, the sufficient statistics  vector $T(g)$ is the degree sequence of the graph $g$. 
 Indeed, the model is usually given directly in the exponential family form~\eqref{eq:expfam} with $T(g)=(d_1,\dots,d_n)$ being the degree sequence of $g$: 
\begin{equation}\label{eqn:beta}
	p_\beta(g) = \exp\{\sum_{i=1}^n d_i\beta_i - \psi(\beta)\}. 
\end{equation}
For a brief history of the model, see for example the introduction and references given in \cite{RPF:11}, which  deals with the  general version of the $\beta$ model and the special case of  the simple graph version  studied in \cite{CDS11}. 
\end{example}

The $\beta$ model~\eqref{eqn:beta} is indeed one of the simplest interpretable undirected random graph models of relevance in statistical applications. In fact, the study of the degree sequences and, in particular, of the degree distributions of real networks is a classical topic in  network analysis, which has received extensive treatment  in the statistics literature 
(starting with, e.g., \cite{HL81}, \cite{FW81}, \cite{fien:meye:wass:1985}), the physics literature (e.g. \cite{NSW:01}, 
\cite{N:03}, \cite{PN:04}, 
 \cite{Will:09}) as well as in the social network literature (e.g., 
 \cite{Robins:08}, \cite{Goodreau}, \cite{HM:07}, and references therein). See also the monograph by  \cite{F-review}  and the books by \cite{Kolaczyk:09} and \cite{Newman:10}. 
 Its known properties  can be used as a blueprint for consideration of more complex models. Specifically, it is a very nice and well-understood ERGM from the following points of view that simultaneously serve as an outline of the remainder of this chapter: 
\begin{enumerate} 
\item Exact inference for model fitting:  
exact testing 
 is used for testing goodness of fit of a model when applicability of standard statistical asymptotic methods is unclear. Typically, this is the case for small sample sizes $N$, and also largely remains a problem for  network data in particular. 
 These topics are the content of Section~\ref{sec:exactTesting}.  
 	\begin{enumerate}
	\item For the $\beta$ model,  exact testing 
	 depends on sampling graphs with a prescribed degree sequence, for which there are several algorithms throughout the statistics, computer science and graph theory literatures. The general problem is wide open for other ERGMs. 
	\item To construct an appropriate Markov chain for exact testing, sampling graphs with fixed properties should be done within the context of a statistical model. 
	\item As we will see, \emph{all} linear ERGMs reduce to hypergraph degree sequence problems. 
	\end{enumerate} 
\item Parameter estimation and noisy data: 
	Viewing random graph models through the lens of exponential families offers a way to capture the issue of existence of maximum likelihood estimators (MLE), a fundamental problem that is largely unexplored. 
	The problem and its statistical implications are explained in Section~\ref{sec:polytopes}.
	\begin{enumerate}
	\item For the $\beta$ model, the geometry of MLE existence is captured by the convex hull of all degree sequences. 
			The extreme points of that polytope correspond to threshold graphs  and facet-defining inequalities have been characterized.  
	In general,  the geometry of  MLE existence  is captured by the model polytope, whose lattice points are realizable sufficient statistics vectors. These polytopes are not known for other ERGMs. 
	\item In data privacy considerations, or in dealing with noisy data, observed sufficient statistics may contain errors and thus in particular may not be realizable by any graph. For the $\beta$ model, it is known when an integer sequence is graphical, i.e., when there exists a graph that realizes the sequence as its degree sequence.  The general problem of characterizing realizable sufficient statistic vectors is wide open yet crucial for establishing reliability of inference. 
	\item Efficient facet description of the model polytope for other examples of ERGMs would be crucial to develop appropriate tools for dealing with noise in the data. 
	\end{enumerate} 
\item The model has desirable asymptotic properties \cite{CDS11}, closely related to graphons \cite{LovaszSzegedy}. As the asymptotics are different than the asymptotics  mentioned in item (1) above, we will not discuss these issues in this Chapter, but instead point the interested reader to \cite{WolfeOlhede13} and for a quick list of references to \cite[Introduction]{RPF:11}.
\end{enumerate}


Here are examples of some other models that build on the $\beta$ graph model and for which many of these questions remain open. In the interest of space and readability,  equations repeating the structure \eqref{eq:expfam} for each case are omitted but can easily be found in the references. 
\begin{example}\label{ex:JDMmodel}
	 The  \emph{joint degree matrix  model} is the ERGM of the form \eqref{eq:expfam} where  the sufficient statistic $T(g)$ is the joint degree matrix, or JDM for short, of $g$. The JDM  counts  the number of edges between nodes of given degrees, for all degree pairs. 
	 The model, as an ERGM, was introduced to the statistics literature in recent work 	  \cite{AleKayvanDKgraphs}, motivated by previous  lines of research outside of mainstream statistics,  such as \cite{ErdosMiklosTorockai14}, which offers a fast mixing algorithm and a fantastic introduction and overview of the joint degree matrix as a network statistic, and \cite{StantonPinarJDM}. 
\end{example}
As pointed out in \cite{ErdosMiklosTorockai14}, fixing the value of the JDM of a network  is stronger than just fixing the degree sequence, though it uniquely defines the degree sequence. 

\begin{example}\label{ex:betaHypModel}
The \emph{$\beta$ model for  hypergraphs} was introduced in \cite{betaHypergraphs}. 
It looks identical to the $\beta$ model for graphs, except the edges in the network $g$ can be of size larger than $2$. In particular, there are three variants: uniform, layered uniform, and general.  
In the uniform variant, hyperedges of fixed size $k$ occur independently with probability proportional to $\beta_{i_1}\cdots\beta_{i_k}$ for all $k$-tuples $i_1,\dots,i_k$ of vertices on the graph. The sufficient statistic in the this case is the hypergraph $k$-degree sequence, that is, the number of edges of size $k$ to which each vertex belongs. 

In the layered variant, the set of possible hyperedges is extended to include edges of all sizes up to and including some fixed size $k$. The sufficient statistics are thus layered hypergraph degree sequences: the number of hyperedges \emph{of each size}, $2,\dots,k$, to which a vertex belongs. 
In the general variant, all edge sizes are allowed, and the model is more complicated; for details see \cite{betaHypergraphs}. 
\end{example}
\begin{example}\label{ex:p1model}
The \emph{$p_1$ model for random graphs} assigns probabilities to \emph{directed} edges in a random graph according to the propensity of the nodes to receive, send, or reciprocate edges. The sufficient statistics vector $T(g)$ of the model consists of: the in-degrees of all nodes, out-degrees of all nodes, and the number of reciprocated edges, where a reciprocated edge is of the form $i\leftrightarrow j$. 
The model has a history in statistic and applications, see for example  \cite{HL81}, \cite{FW81} and \cite{PRF10}. 
The last reference interprets the model through the lens of algebraic statistics and studies its linear ERGM structure in terms of algebra and geometry; cf.\ \cite{RPF:11}. 
\end{example}
A fairly recent extended review article on  statistical network models  \cite{F-review}  offers many other examples of models, beyond linear ERGMs, whose geometric and combinatorial properties are yet to be explored.

\section{Sampling algorithms in exact testing}\label{sec:exactTesting}

 Testing the fit of the model means  deciding whether the model provides a plausible probabilistic representation of the available data.
 Unfortunately, most tests for goodness of fit are based on large sample approximations that are not applicable to data such as  sparse contingency tables used in, for example, cross-classification of categorical variables, or data such as random graphs  that can be naturally  thought of as contingency tables through incidence matrices. 
Fundamental examples of inadequacy of the use of asymptotic approximations were already pointed out over two decades ago in  statistics literature \cite{Haberman88}, \cite{Agresti}: when sample size $N$ is small, or for example if some data table cell entries are much smaller than others, 
  \emph{exact testing} should be performed instead. 
Even so, a large part of the literature on, say, network computation and modeling does not address the lack of model fitting and testing methodology\footnote{The reader is urged to distinguish between this statistical questions of model fitting (`is the model \emph{correct}?') and parameter estimation ('assuming the model is correct, find the parameter value(s) that best fit the data'). The latter is often alluded to within the context of `model validation' in the computer science literature.} beyond heuristic algorithms  \cite{H:03, NtwkApproxFittingDynamics15, hunter2008goodness}. This is largely due to the inherent model complexity or degeneracy and the lack of tools that can handle network models and sparse small-sample data. 

The role and relevance of discrete mathematics 
 in exact testing is as follows.  
In an exact test for a model with sufficient statistics vector $T$, the observed data  $x$ with    $T(x)=t_{obs}$ is compared to a reference set  \[\mathcal F_{t_{obs}}:=\{y:T(y)=t_{obs}\}\] of data with the same value of the sufficient statistic. This reference set is also called the \emph{fiber} in algebraic statistics literature. It is a fiber of the  algebraic map that computes the sufficient statistics; in the $\beta$ model case the map is linear, since node degrees can be computed as row/column sums of the incidence matrix.  Note that, by definition of sufficiency, the probability of any data point in the fiber is determined completely by the value $t_{obs}$, hence the fiber is a good reference set to consider for testing. 
Using a suitable choice of a goodness-of-fit statistic that measures, for example, the distance of the data $x$ from the expected value under the model, the simulated data points in  $\mathcal F_{t_{obs}}$ are compared to the observed data. If a relatively large number of those are closer to the expected value than the observed data, than one declares  poor model fit. 
 These are, roughly,  the essential elements of an exact test, whose name refers to the exact conditional distribution of data points in the fiber $\mathcal F_{t_{obs}}$, where we are conditioning on the observed value $t_{obs}$ of the sufficient statistics.  

\begin{figure}
	\includegraphics[scale=0.7]{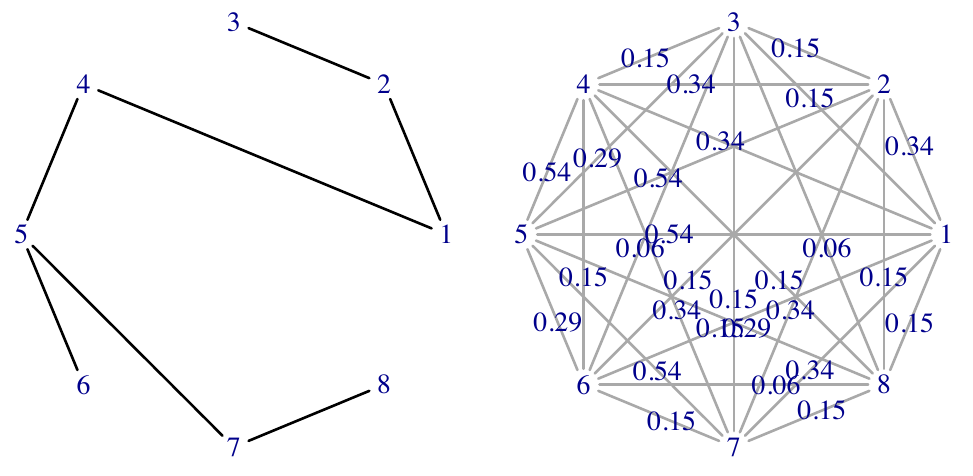}
\caption{Observed graph $g_1$ (left) with degree sequence (2,1,1,2,3,1,2,1), and the expected probability-weighted graph under the $\beta$ model.}
\label{fig:plotGraphAndMLE}
	\includegraphics[scale=0.6]{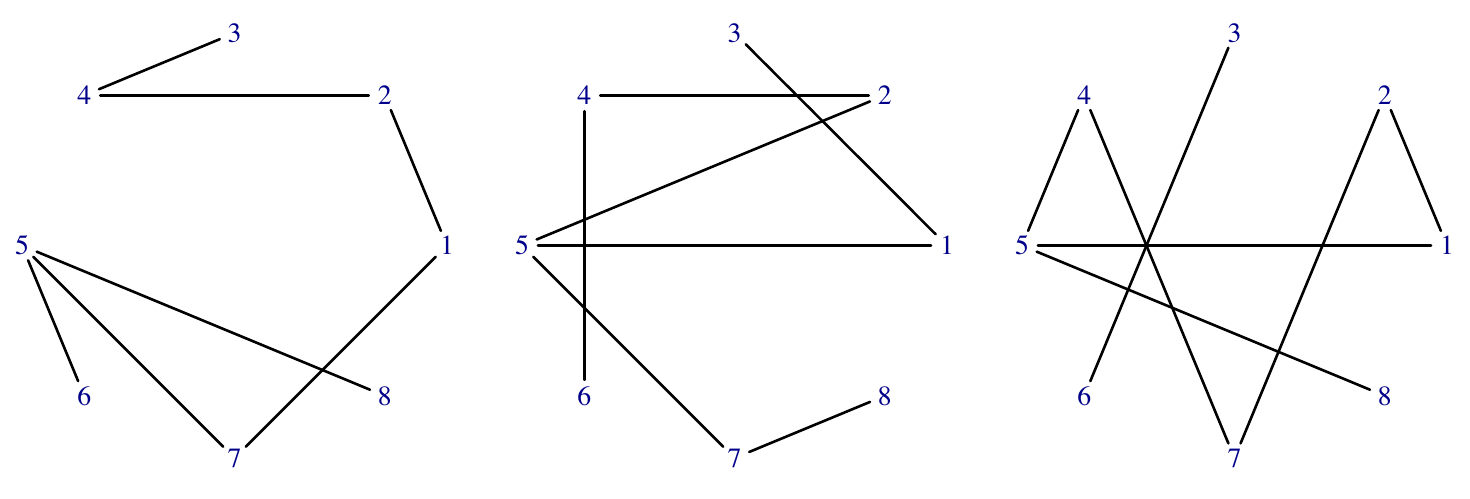}
\caption{Three out of the $590$ other graphs in the fiber $\mathcal F_{(2,1,1,2,3,1,2,1)}$: $g_2$ (left), $g_3$ (center), $g_4$. }
\label{fig:Plot3graphsInFiber}
\end{figure}

To visualize the elements of an exact conditional test, consider the following example. 
Suppose the observed graph is $g_1$ in Figure~\ref{fig:plotGraphAndMLE}; see \cite[Figure 13]{OHT:11} and \cite[Section 4.1]{GPS14}. Given the degree sequence of $g_1$,  the expected value under the model is visualized as a weighted graph whose edge weights are the expected values of the random variables representing the edges. This is computed using the MLE of the parameters; see Equation~\eqref{eq:MLE}. One may view these expected values as computing the relative frequency of the edges in the set of all graphs whose  degree sequence equals that of the graph $g_1$. For example, the edge $\{3,4\}$ appears in 15\% of graphs that realize this degree sequence, and since the conditional distribution on the fibers of the $\beta$ model is uniform, this frequency doesn't need to be weighted to compute the expected values. 
Intuitively, the exact test is meant to assesses the relative closeness of the observed  graph  to the expected graph, and thus it answers the question: is $g_1$ more like the expected graph than the graphs $g_2, g_3, g_4$ from Figure~\ref{fig:Plot3graphsInFiber} and the other $587$ graphs in the fiber? 

 There are two important points to be made before we discuss the key problem. First, choosing a good goodness-of-fit statistic is in general an open problem, and is difficult in particular in random graph models for which edges are not independent random variables. For a discussion of  the special case of  2-way contingency table models, see \cite[\S 3.1]{Agresti}. Second, a good example of a goodness-of-fit statistics to use for random graphs with independent edges is the chi-square statistic, which simply computes the square of the weighted  Euclidean distance in $\mathbb R^n$ between the observed graph and the expected graph (or MLE). 
  It is defined as: 
 \[
 	\chi^2(g):=\sum_{i,j}\frac{(\hat{g}_{ij}-g_{ij})^2}{\hat{g}_{ij}}, 
 \]
 where $g_{ij}$ is $1$ if the edge $\{i,j\}$ is present in $g$ and $0$ otherwise, and $\hat{g}_{ij}\in(0,1)$ is the expected value of the random variable representing the edge $\{i,j\}$ under the MLE, as illustrated in Figure~\ref{fig:plotGraphAndMLE}.
 The values of the chi-square statistic for the four graphs above are: 
 \[
 	\chi^2(g_1)=19.49 , \quad
 	\chi^2(g_2)=17.42, \quad 
 	\chi^2(g_3)=21.56, \quad
 	\chi^2(g_4)= 26.53.
 \]
 To determine whether  the value $\chi^2(g_1)$ is too large, one should enumerate the fiber and compute the value of the statistic for all $591$ graphs in it.  
 
As it is computationally prohibitive to enumerate the fiber except in very small cases, one tries to instead sample from the fiber and estimate the number of data points that are more extreme than the observed, with respect to the  chosen goodness-of-fit statistic.    
Thus the {\bf key problem} in the exact testing procedure is that of exploring the fiber through sampling from the conditional distribution  given $t_{obs}$. 
\begin{problem}[General Problem 1]\label{prob:GeneralSampling}
	Fix a random graph model $\mathcal M$ with sufficient statistics  $T$. Develop an efficient algorithm that samples from the space of graphs with arbitrary fixed value of $T(g)=t_{obs}$. 
\end{problem}

A fiber sampling algorithm can be a direct sampler or it can be based on a Markov chain and used  through the Metropolis algorithm, which is a standard algorithm and can be found, for example, in  \cite{RC99}.  The algebraic statistics literature focuses on the latter. The reason for this is the natural connection to algebraic geometry. Namely, 
the theoretical feasibility of constructing Markov chains to solve  General Problem 1 for arbitrary \emph{linear} ERGMs, as well as  finiteness of the complexity  of steps   needed to perform the Markov chain random walk on fibers of \emph{every} log-linear model, form a cornerstone of traditional algebraic statistics through a fundamental theorem from \cite{DS98}; see also \cite[Theorem 1.3.6]{DSS09}. 
Its applied solution and relevance in exact testing for any linear ERGM 
 hinges upon  the development of efficient sampling algorithms for graphs and hypergraphs, as outlined here. 
 
 The steps or moves needed to perform the Markov chain above for a a given model are called \emph{Markov moves}. A \emph{Markov basis} for a model is a collection of moves guaranteed to connect each fiber of that model. 
A move is an element of the \emph{Graver basis} of the model if it contains no proper sub-moves; see Figure~\ref{fig:Plot1moveOnFiber} for a running example. 
 
\begin{remark}
In order to be of statistical relevance, and make any inference from the observed data, Problem~\ref{prob:GeneralSampling} should be solved within the context of a statistical model.  In other words, when a network feature, or a set thereof, is fixed in sampling, it is understood  that the feature can serve as a sufficient statistic of some model; the statisticians then ask to identify the model and study its various properties in order to determine feasibility and reliability of inference and model fitting.  
\end{remark} 
\begin{example}\label{ex:betaModelFiber}
	For the $\beta$ model for random graphs from Example~\ref{ex:betaModel}, the reference set $\mathcal  F_{t_{obs}}$ is the set of graphs whose degree sequence $d$ equals the degree sequence $t_{obs}:=d(g)$ of the observed graph $g$.  
	It is easy to see that in the $\beta$ model the conditional probability distribution of graphs given  a fixed observed degree sequence is uniform; in general, of course, this distribution depends on the model. 
Therefore, in the $\beta$ model, as well as others we discuss in this chapter as the reader may easily verify, graphs should be sampled uniformly, because conditioning on the sufficient statistics results in the uniform distribution on the fiber.  
\end{example}

Of course, Problem~\ref{prob:GeneralSampling} has been solved for the $\beta$ model for graphs.  
Indeed, various algorithms for constructing random graphs with a fixed degree sequence - in other words, sampling the fibers of the $\beta$ model - have been proposed over the years, with or without the context of the model,  and the reader is probably familiar with the graph theory literature on this topic, most relevant part of which is cited throughout this Chapter. 
On the statistics side, \cite{BlitzDiac10} derive a  sequential importance sampling algorithm for this model.  A different random generation algorithm for simple connected graphs can be found in  \cite{VigLat05}. 

As mentioned above, one could think of solving the sampling problem in different ways:  by an algorithm that  constructs, uniformly at random,  a graph that realizes the given degree sequence (where care should be taken that such an algorithm can, in fact, discover the entire fiber!); 
or by  finding a set of moves, sometimes also called  `edge swaps' in graph theory, that can serve as a basis for a Markov chain Monte Carlo (MCMC) sampling algorithm on the fiber. Here the word `basis' is used in a technical sense but can also be understood intuitively: the moves should guarantee to discover the entire fiber and define a symmetric, aperiodic chain on \emph{any} given fiber of the model. 
The Markov chain approach  offers a nice alternative and an interesting  connection to other areas of mathematics. As such, it is largely the focus of the remainder of this section. 

\subsection{Ingredients for a Markov chain fiber sampler} 
In order to use the MCMC method to sample the fiber of an observed graph, three important questions should be answered. First, a \emph{Markov basis} should be specified: a set of moves that the algorithm uses to go from any graph to any other graph in any fiber. Second, the \emph{stationary distribution} of the proposed Markov chain should be the correct conditional distribution; recall that in the $\beta$ model example, it should be uniform. Third, \emph{mixing time} considerations cannot be avoided in this approach - one can easily propose seemingly straightforward Markov chains that have extremely unreasonable behavior in that they will take a very long time to converge or to explore the fiber.  

One way to answer the first two questions  simultaneously is through the use of algebraic statistics. The third question on mixing time  is discussed subsequently. 

\begin{figure}
	\includegraphics[scale=0.5]{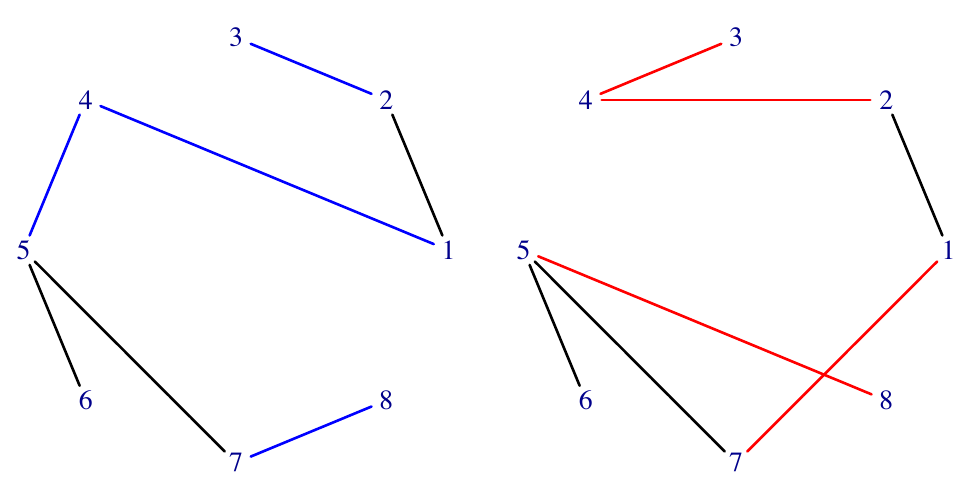}
\caption{A Markov  move under the $\beta$ model for the graph $g=g_1$ from Figure~\ref{fig:plotGraphAndMLE}: the graph $g_1$ with edges $h_2$ to be removed in blue (left), and the resulting graph $g_2$ from Figure~\ref{fig:Plot3graphsInFiber} after $h_2$ is replaced by edges in $h_1$ which is  shown in red. Note that this move $h_1-h_2$ is not a minimal one, cf.\ Figure~\ref{PlotSeqQuadMoves}. Instead,  it is an example of a Graver move. If structural zeros were present in the model, moves like this one may be needed for connecting the fiber.}
\label{fig:Plot1moveOnFiber}
\end{figure}

There is an algebraic geometry or, if you will,  commutative algebra equivalent to a Markov basis for a linear ERGM stemming from the fundamental theorem from  \cite{DS98}. Namely, every such statistical model  lies in a natural algebraic variety: A \emph{variety} is the set of all solutions to a system of polynomial equations, and those encoding parametric statistical models can equivalently be described parametrically -- there is an algebra-geometry duality here; see the excellent introductory explanation in \cite[\S 2.4 and 2.5]{KarenInviteAG}.  Focusing on the algebra, the parametrization is encoded via a homomorphism between two polynomial rings, say $\phi:\mathbb C[E]\rightarrow \mathbb C[\theta_1,\dots,\theta_n]$, where $E$ is some set of  variables, and $\theta_i$ are parameters. 
 The indeterminates in the two polynomial rings carry a special meaning in algebraic statistics: $E$ represent random variables of interest, while $\theta=(\theta_1,\dots,\theta_n)$ is the vector of unknown  model parameters. For example,   in the case of ERGMs whose edges are independent random variables, $E=E(G)$ are edges of the random graph $G$. 
The kernel of the map $\phi$ is an ideal in the polynomial ring $\mathbb C[E]$, and this ideal is the defining ideal of the algebraic variety whose real positive part contains the statistical model. 
Readers interested in further algebraic  details of this correspondence may consult \cite[page 25]{DSS09}. 
 \begin{example}\label{ex:toricBeta}
 In the specific case of the $\beta$ model for graphs  with no statistical sampling constraints, the graph  $G$ is the complete graph  because all edges are allowed to appear in the random graph realization $g$. The homomorphism $\phi$ maps each edge $e_{ij}$  to the product of the parameters $\beta_i\beta_j$. 
For example, consider the random graph $\beta$ model on $n=8$ nodes. The coordinate map   
 corresponding to the parameterization of the model is 
\begin{align*}
 \phi_{K_8}:\mathbb C[e_{12},e_{13},\dots,e_{78}] &\rightarrow \mathbb C[\beta_1,\dots,\beta_8]\\
 		e_{ij}&\mapsto \beta_i\beta_j.
\end{align*}
 Note that here we have forgotten the exponential and the normalizing constant because they do not affect the algebraic structure of the parametrization, so by abuse of notation we drop the `$\exp$' from the model specification. 
An example of an equation that vanishes on the model is $e_{17}e_{24}e_{34}e_{58}-e_{14}e_{23}e_{45}e_{78}$. The algebraic relation holds because $\phi_{K_8}(e_{17}e_{24}e_{34}e_{58})=\phi_{K_8}(e_{14}e_{23}e_{45}e_{78})$ since, of course,  the two graphs on edge sets  $\{17,24,34,58\}$ and $\{14,23,45,78\}$  have the same degree sequence, which is the sufficient statistic for the model; refer to Figure~\ref{fig:Plot1moveOnFiber}.  Note that the equation vanishing on the model  is equivalent to  saying that equation is in the kernel of the coordinate map: $e_{17}e_{24}e_{34}e_{58}-e_{14}e_{23}e_{45}e_{78}\in\ker\phi_{K_8}$.
 \end{example}
To summarize: probability distributions in the model correspond to real positive points on an algebraic variety, where the parametric description of the model gives rise to the parametrization of the variety. This variety is  defined  implicitly by the equations that vanish on the model. 
Coordinates of the  polynomial ring that is the domain of $\phi$ correspond to the joint probabilities of random variables in the model. Equations that vanish on the variety represent algebraic relations among the joint probabilities. To visualize what these relations mean, consider  $\prod_{\{i,j\}\in E^+} e_{ij}- \prod_{\{i,j\}\in E^-} e_{ij}$, for two sets of edges $E^+$ and $E^-$. This  is an equation that vanishes for all points on the model (i.e., a relation on the edges) if and only if the graphs whose edges sets are $E^+$ and $E^-$ have the same probability under the model.  In turn this happens if and only if the two subgraphs have the same values of the sufficient statistics vector.
	In the linear ERGM case, the variety has a special structure called toric, due to linearity,   and as a consequence  the defining equations are all binomials.
	
	The crucial observation is then that such binomials correspond to moves on the fibers of the model, and in fact \emph{the moves comprise a Markov basis if and only if the binomials suffice to generate all equations that vanish on the variety} \cite{DS98}. This fact is sometimes called the fundamental Theorem of Markov Bases. This correspondence  should be  interpreted in the following way: $h_1-h_2$  is a defining equation of the variety corresponding to the model if and only if replacing the subgraph $h_2\subset g$ by the subgraph $h_1\subset g$ is a move on the fiber of the model, meaning that the move does not change the value of the sufficient statistics vector, as illustrated in Figure~\ref{fig:Plot1moveOnFiber}. 

In fact, the correspondence applies to general log-linear models for discrete data, not only networks, and is one of the cornerstones of algebraic statistics. In the general case the moves $h_1-h_2$ need not be squarefree and need not represent graphs; instead each $h_1$ and $h_2$ can be an arbitrary monomial in the indeterminates $E$ and this machinery works for general contingency table models. Simple graphs are just 0/1 contingency tables, while graphs can in general be represented by non-negative tables, where each positive entry indicates presence of an edge, with multiplicity if the model allows for it. 
  Given the very extensive literature on this topic, for example summarized in the recent book \cite{AHT2012} and references given therein, let us not spend more time  studying details of  this equivalence here. Instead, we pause to note the important consequences: 
 By a fundamental theorem in commutative algebra called the Hilbert basis theorem, a Markov basis exists and is finite for \emph{every} linear ERGM. Furthermore, using the moves from a Markov basis given by the algebraic correspondence theorem  to generate moves in the Metropolis algorithm \emph{automatically} converges to the correct stationary distribution on the fiber.  Note that, by definition,  any set of moves containing a minimal Markov basis has the same property.

The $\beta$ model for  graphs from Example~\ref{ex:betaModelFiber} somehow seems to have a parallel history in various literature, as reflected upon in the following Remark. 
\begin{remark}\label{rmk:betaModelSamplingLiterature}
 In commutative algebra, Markov bases for the $\beta$ model exist under the name of \emph{generators of  toric ideals of graphs}.  This name comes from the fact that the moves from the $\beta$ model 
 	 can be encoded by \emph{closed alternating walks on graphs}: sets of edges  partitioned into two sets, $E^+$ and $E^-$, such that they constitute a closed walk when traversed in alternating order $e_1,\dots,e_m$ with $e_i\in E^+$ for even $i$ and $e_i\in E^-$ for odd $i$.  	 
 For example, the move corresponding to removing blue and adding red  edges in Figure~\ref{fig:Plot1moveOnFiber} represents one such alternating walk, with $E^+$ and $E^-$ corresponding to red and blue edges, respectively.  
The reader will note that this move can be obtained as a sequence of three moves represented by closed 4-cycles illustrated in Figure~\ref{PlotSeqQuadMoves}. Equivalently, the polynomial $e_{17}e_{24}e_{34}e_{58}-e_{14}e_{23}e_{45}e_{78}$ lies in the ideal generated  by the three polynomials representing the 4-cycles: $e_{17}e_{24}e_{34}e_{58}-e_{14}e_{23}e_{45}e_{78} = e_{24}e_{34}(e_{17}e_{58}-e_{15}e_{78})+e_{45}e_{78}(e_{13}e_{24}-e_{14}e_{23})+e_{24}e_{78}(e_{15}e_{34}-e_{13}e_{45})$.
\begin{figure}
	\includegraphics[scale=0.5]{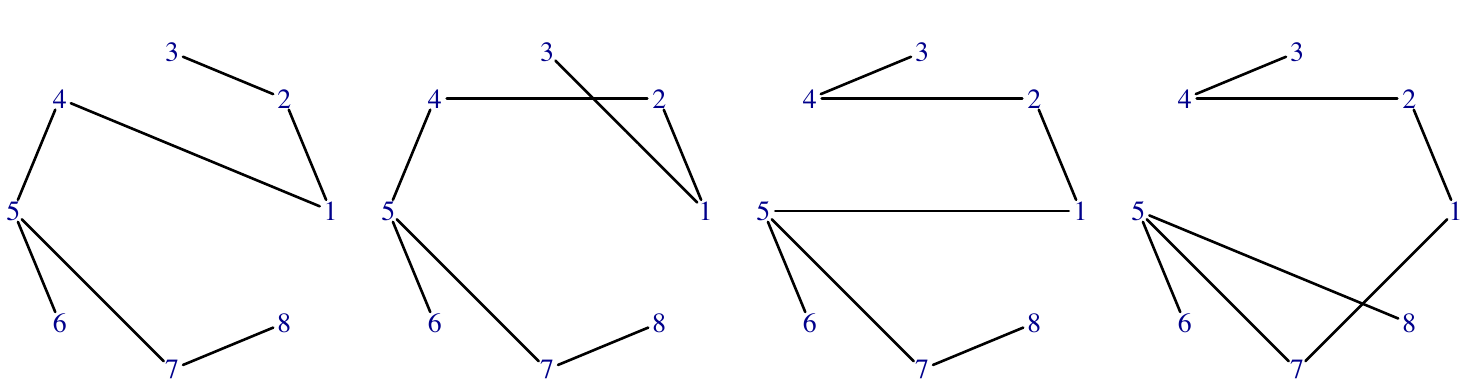}
\caption{A walk using only quadratic moves from $g_1$ to $g_2$ on the fiber $\mathcal F_{(2,1,1,2,3,1,2,1)}$.}
\label{PlotSeqQuadMoves}
\end{figure}

The commutative algebra community has known the structure of these generators, without calling them moves or relating them to MCMC methods, for at least 20 years, for example, in  \cite{Vill95, Vill} and \cite{OHquad99}, as well as  \cite[Chapter 9]{St} and \cite{RTT12}.  

Further,  these results  have been known in the algebraic statistics literature for some time as well. Notable examples include works  such as \cite[Theorem 4]{SesaSethToric} on toric geometry of compatible full conditionals, \cite[Theorem 3.2]{Morton10} on relations among conditional probabilities, \cite{StWe12} that studies algebra and geometry of statistical ranking models and the ascending model in particular, and \cite{PRF10} that study the structure of Markov and Graver bases for the directed model $p_1$  whose restriction (to the  part of the random graph consisting only of reciprocated edges) is the $\beta$ model. More recently, \cite{OHT:11} offer the exact test implementation whose main ingredient is  a random generation of Markov and Graver bases elements for the $\beta$ model. 
  \end{remark} 
Interestingly, bicolored or alternating closed walks on graphs make a later appearance in computer science literature as well.  \cite{ConeBalancedGraphs:AmitPeledSrini:09} define the \emph{alternating cone} of a graph to be the set of characteristic vectors of alternating walks on the graph, cf.\ \cite{ConeAlterWalks:AmitPeledSrini:07} and  the subsequent book chapter \cite{Amitava:BookChp:ConesWalks}. In algebraic terms, this is the cone of exponent vectors of the binomials in the toric ideal of that graph. 
Results from toric algebra of graphs and the definition of the alternating cone  should easily imply that \cite[Conjecture 3.1]{ConeBalancedGraphs:AmitPeledSrini:09} is actually true, and can furthermore be generalized for hypergraphs appropriately.  \cite{SriniProblems} is a  very nice set of notes on open problems related to threshold graphs; cf.\ Conjecture 4. To prove the conjecture, note that 
a `balanced sum of cycles' is certainly a balanced graph, thus it is a lattice point in the cone of balanced subgraphs.  Each of these lattice points is an integer exponent of some binomials in the toric ideal of the graph. It is well known that the Graver basis is an integral generating set of the lattice, in that every lattice point in the cone can be written as a integral sum of the Graver exponents. 
 Finally,   Graver bases of these cones for graphs are encoded by alternating  walks by \cite{Vill95}, see also recent literature cited above. More generally, for hypergraphs, the walks are interpreted as bi-colored \cite{PS} and again form a Graver basis; see  \cite{bouquetAlgebra} for the general non-uniform case.  By definition, these are exactly balanced subgraphs. Thus, the exponent vector of a  \emph{balanced sum of cycles is an integral sum of balanced subgraphs} represented by Graver elements. 
The present author is not aware of whether the conjecture has been proved in other literature;  only the \emph{fractional} sum part of the conjecture is proved in the  works  on alternating cones cited above. 
\smallskip

Apart from the $\beta$ model for random graphs, sampling algorithms for the space of graphs with fixed properties have evolved disjointly from statistics. 
At this point,  solutions to the following two subproblems of Problem~\ref{prob:GeneralSampling} are of special importance: 
\begin{problem}[An instance of General Problem 1: $p_1$ model moves]\label{prob:p1moves}
	Determine a set of moves sufficient to connect the space of graphs, consisting of both directed and bidirected (reciprocated) edges, with a given directed  degree sequence and number of reciprocated edges. 
\end{problem}
The problem above was solved in algebraic statistics \cite{PRF10} and an implementation of a dynamic algorithm was offered in \cite{GPS14}; see  the subsection on mixing times below for a related open problem. 
\begin{problem}[An instance of General Problem 1: Hypergraph $\beta$  model moves]\label{prob:BetaHypSampling}
	Determine a set of moves sufficient to connect the space of hypergraphs,  uniform, layered or general, with a given  degree sequence. 
\end{problem}
Of course, one can `solve' these problems in various ways. From the point of view of (algebraic) statistics applied  to networks, the most interesting types of solutions to Problems~\ref{prob:BetaHypSampling} and \ref{prob:GeneralSampling} in general will \emph{at least address} general models with sampling constraints and forbidden edges, or derive  bases of bounded complexity with good mixing time properties for a subclass of the models. The literature offers some examples of such solutions, but many problems remain open.  These issues are discussed next. 

\subsection*{Sampling constraints} 
If, in applications, the random sampler of the model fibers is allowed to step through the extended state space of  non-simple graphs, where edges can have a multiplicity, then the problem is solved. For example, simple-edge-swap solution to Problem~\ref{prob:BetaHypSampling} follows from standard textbook results in algebraic geometry; namely, the generating set of the toric ideal for the model is known to consist of quadratic binomials, as the algebraic closure of the model is 
defined by the toric ideal of the second hypersimplex, known to be generated by quadrics. 
However, inherent sampling constraints can compound the issue of computing a set of moves guaranteed to connect each fiber, for example, the sampler may need to step through only simple graphs, or graphs with bounded multiplicity. In some statistical modeling considerations other constraints may come in the form of forbidden edges or forbidden neighbors; such constraints in statistics are called \emph{structural zeros} of the model. 
 Of course sampling and modeling  constraints restrict the fiber, as expected, but also make the required moves more complicated: consider the situation when the edges used in the intermediate steps of the walk in Figure~\ref{PlotSeqQuadMoves}, such as the edge $\{1,3\}$ for example, are structural zeros. 
 
 Interestingly, \cite{ToroEtAl:degConstruction} characterize when a sequence is realizable as a degree sequence of a simple graph such that a given set of edges from an arbitrary node is avoided. This allows the authors to construct a swap-free algorithm for sampling the restricted fiber of the $\beta$ model. 
 If one is interested in sampling a restricted fiber for any general linear ERGM or any log-linear model, \cite{OHT:11} and \cite{AHT2012} show that one needs a much larger set of moves to guarantee connectivity, i.e.\ minimal Markov bases 
  may not do the job. For example, the set known to algebraists to be inherently more complicated, called the Graver basis of the toric ideal \cite[\S 1.3]{DSS09}, \cite[\S 4.6]{AHT2012} will suffice to connect the fiber, however it is also so complex and large that it is not feasible to compute it for any reasonable example beyond toy-size networks; to give an idea of such size, for the $p_1$ model this basis can only be directly computed for networks on less than 7 nodes. 
 What \cite{OHT:11} then do is provide an algorithm that constructs one step in the chain -- a Markov move -- in a dynamic fashion and show its performance on several small network datasets. Since the moves generated belong to the Graver basis, the method  is  provably applicable in the case of structural zeros and sampling constraints. 
This method does rely on  the understanding of the algebraic machinery underlying  the model, but appears to be inefficient in discovering the fiber and the resulting Markov chain has a high rejection rate. In contrast, simulations indicate that the chain from \cite[\S 4.1]{GPS14} does not have that problem, but there is no formal proof of its good properties.
This leads us to the next topic regarding sampling fibers. 

\subsection*{Mixing time considerations}
Considerations of mixing times/convergence are crucial in any MCMC sampling scheme \cite{MCMCmethodsBOOKwithR} \cite{MCMCmethodsHandbook}. 
Notably, using minimal Markov bases may not produce good mixing as seen in some experiments and pointed out recently in \cite{MBrapidmixing}, which formalizes how mixing properties of Markov bases depend on the structure and connectedness in particular of the underlying \emph{fiber graph}. Moreover, it explains that results on mixing times of  chains built on minimal Markov bases are in general prohibitive to obtain, and  suggests instead the construction of related expander graphs for the problem - a worthwhile pursuit. 
In earlier work,  \cite{GPS14}  construct a chain on the fiber graph for  the $p_1$ model in  in such a way that it is, in fact, a complete graph, with loops having seemingly low probability. In simulations, the chain behaves quite well and indicates fast mixing.  Nevertheless, a formal proof of rapid mixing or any  mixing properties are left as  open questions, as the authors there did not study the structure of the transition matrix for that chain. 

This is where recent advances in discrete  literature, for some reason disjoint from statistics literature on networks, offer  example results for  the $\beta$ model that, in this author's opinion,  should be connected to statistics - algebraic statistics in particular -  and used as a guide for studying Markov bases of other linear and general  ERGMs. 
Namely, \cite{KannanEtAl99} conjectured that the Markov chain based on swaps mixes rapidly, and  
 the proofs for arbitrary regular graphs and half-regular bipartite graphs  were given in  \cite{CooperDyerGreenhill07} and \cite{MiklosErdosSoukup13}, respectively, while irregular graphs are studied in \cite{Greenhill15}. 
Recently, \cite{ErdosMiklosTorockai14} offer insight to the general case via the study of swap-based chains on the fibers of the JDM model from Example~\ref{ex:JDMmodel}; they prove fast mixing on a subset of the JDM-model fibers and offer an excellent review of related literature, while irreducibility of the Markov chain on the JDM realizations was proved in \cite{CzDErdMikl5}. 
To translate what this means for Markov bases, note that 
 in algebraic statistics language, `swaps' of two edges at a time are  quadratic Markov moves, or quadratic generators of the toric ideal, since taking two edges at a time and replacing them by two other edges translates to a quadratic monomial in Example~\ref{ex:toricBeta}.  
 
Crucially, however, the difference  is not just linguistic:  in general in algebraic statistics, Markov moves are sampled  from a set of all possible moves that apply to the model, that have been computed a priori. Such sampling  can  clearly lead to more rejections. On the other hand, `swaps' are constructed from observed edges, so rarely is a non-applicable swap proposed! 
In particular, the literature cited here provides a set of examples of  \emph{rapidly mixing sequences of fiber graphs}, and thus also hope that Markov-move-based constructions, adapted appropriately, can still lead to good chains for fiber exploration and goodness-of-fit tests; compare to examples reported  in \cite{GPS14} or \cite{OHT:11} for an illustration. 

In the remainder of this subsection, let us clarify how  these chains should be constructed `with care', and why an out-of-the-box Markov basis may not be good enough. 
 For example, \cite{MBrapidmixing} observes  that enlarging minimal Markov bases to, say, Graver bases doesn't improve mixing time.  Intuitively, if the chain can be constructed so that the fiber graph has a relatively small number of loops, then rapid mixing may be possible. Such a fiber graph would resemble the idea of using moves corresponding to graph-theoretic `swaps':  namely, not attempting to construct or apply Markov moves that work on just some fiber, but, instead, focus on the given observed graph and its particular fiber. We refer to such moves as \emph{data-dependent}. 
 Incidentally, the chain on the $p_1$ model used  a large set of moves that is guaranteed to contain the Graver bases \emph{in combination with} constructing data-dependent moves  dynamically, involving both `large' and `small' steps on the fibers of the $p_1$ and $\beta$ models,  shows excellent mixing properties in simulations on networks of various sizes. In fact, the Graver move from Figure~\ref{fig:Plot1moveOnFiber} was constructed using the particular implementation from \cite{GPS14}, without going through the intermediate steps in Figure~\ref{PlotSeqQuadMoves}. 
  In the language of Example~\ref{rmk:betaModelSamplingLiterature}, what we construct  is a random walk on the realizable lattice points of the alternating cone 
using a superset of the Graver basis  for the toric ideal of the complete graph, or for an arbitrary graph whose missing edges represent structural zeros in the $p_1$ model. 
This leads us to the final remark about Markov bases construction, which is intimately related to mixing time, but deserves to be singled out. 

\subsection*{Data-dependent samplers}  
That knowing an entire Markov basis for a model may still not be sufficient to run goodness-of-fit tests efficiently is a well-known fact in algebraic statistics. Namely, Markov bases are data-independent \cite[Problem 5.5]{DobraEtAl-IMA}. 
 To paraphrase \cite{AHT2012}: since a Markov basis is common for every fiber, the set of moves connecting the particular fiber of the observed data    will usually be significantly smaller than the entire basis for the model. 
 The subsequent paper \cite{MBrapidmixing} offers another intuitive explanation of why data-dependent fiber samplers are necessary: ``The conclusions we draw [...] are that an adaptation of the Markov basis has to take place depending on the right-hand side", meaning that distinct fibers require different adaptations, confirming previous observations. 
 To handle precisely this issue, \cite{Dob2012} suggested generating only moves needed to complete one step of the random walk, and then covering the fiber by sets of local moves; this was the first  scalable method for exact tests on tables beyond decomposable models \cite{DS04} 
 and applicable  to log-linear models where sufficient statistics are table marginals. However, many models of interest are not captured by marginals \cite{SlaZhuPet, HL81, shellERGM, F-review}, leaving a gaping hole in the methodology for other categorical models, particularly for sparse data. 
 This leads us to consider more general models for networks, with a long-term aim to design   a data-dependent sampler beyond the one in  \cite{GPS14} and \cite{OHT:11}  for the $p_1$ and $\beta$ models.

\subsection{General linear ERGMs} 
The situation may now seem hopeless for  general non-degree-based models, because even if we establish fast mixing for Markov chains on $\beta$ model fibers, there are infinitely many other models that are left to be considered. 
However, not all is lost: it turns out that  as long as the model is log-linear and the sufficient statistics are given by a linear map from the sample space, sampling hypergraphs with prescribed degree sequences, uniform or not, and with prescribed forbidden edges actually suffices for performing the exact test.
This takes  care of a  number of interesting models! 
The answer lies in the structure all log-linear models including linear ERGMs: each of these models is encoded by a hypergraph, defined in \cite{PS}, Section 3.1 of which provides further details and a simple example for the independence model. 
\begin{definition}
The \emph{parameter hypergraph} $\mathcal H_\mathcal M$ of any log-linear model $\mathcal M$ for discrete data, and  any linear ERGM in particular, is the hypergraph whose  vertex  set $\theta_1,\dots,\theta_n$ 
 corresponds to the parameters of the statistical model, and whose edge set is determined by joint probabilities of all possible states  of the random variables $Z_1,\dots,Z_m$.  
More precisely, $\{\theta_j\}_{j\in J}$ 
  is an edge in the parameter hypergraph if the index set $J$ describes one of the probabilities in the model, that is, there exist values $i_1,\dots,i_m$ such that $Prob(Z_1=i_1,\dots,Z_m=i_m)=\prod_{j\in J}\theta_j$.  
\end{definition}
\begin{example}\label{ex:paramHypBeta}
Let $\mathcal M_\beta$ be the  $\beta$ model for random graphs on $n$ vertices $v_1,\dots,v_n$. The parameter hypergraph $\mathcal M_\beta$ is the complete graph on vertices $\beta_1,\dots,\beta_n$. To see why, recall the probability formula of the edges in Example~\ref{ex:betaModel}: 
the random variables $Z_1,\dots,Z_m$ represent the $m={n\choose 2}$ edges of the graph. The probability of the joint state where $Z_k=1$ and $Z_l=0$ for $l\neq k$, that is, the event of  the occurrence of only the edge $Z_k:=\{i,j\}$ in the graph, is proportional to $\beta_i\beta_j$.
\footnote{Of course we take the minimal presentation here; knowing $Prob(Z_1=1,Z_2=0)$ and $Prob(Z_1=0,Z_2=1)$,  we need not consider the joint probability $Prob(Z_1=1,Z_2=1)$.}
Note that  $\mathcal H_\mathcal M$ being complete does not mean that all edges $\{v_iv_j\}$ in the random graph are observed, but instead that all edges $\{\beta_i\beta_j\}$  of $\mathcal M_\beta$ correspond to random graph edges that are \emph{allowed}. In particular this is why the graph on the right of Figure~\ref{fig:plotGraphAndMLE} is complete with non-zero probability weights on every edge- there are no structural zeros in that example. 

Similarly,  for the hypergraph $\beta$ model, the parameter hypergraph is the complete hypergraph, either uniform, layered, or general, respectively in each of the there cases of the model. Recently, the hypergraph $\mathcal H_{p_1}$ of the $p_1$ random graph model  was described in \cite{GPS14} and consists of  edges of sizes 1, 3 and 7, for  probabilities of no connection between two nodes, or a directed edge, or a reciprocated edge, respectively; for example,   a directed edge $i\rightarrow j$ occurs with probability $\lambda_{ij}\alpha_i\beta_j$ and is thus encoded by a $3$-edge on $\mathcal H_{p_1}$.
\end{example}
\begin{remark}
The parameter hypergraph runs the danger of being lost in translation, so to 
 understand its construction, recall that 
the sufficient statistics of $\mathcal M$ are computed using a linear map. Let $M$ be the matrix of that linear map. \emph{The parameter hypergraph $\mathcal H_\mathcal M$ of the model is simply the hypergraph whose  vertex-edge incidence matrix is this matrix $M$.}  \\
For the $\beta$ model, the definition may appear trivialized, but further log-linear model examples show this isn't the case in general; see for example Figure 2 in \cite{GPS14} for two further  examples of parameter hypergraphs. 
\end{remark}

What about the fibers of $\mathcal M$? 
Consider again the  $\beta$ model for graphs: two random graphs $g_1$ and $g_2$ occur with the same probability and are in the same fiber of the model if and only if their images under the linear map $M$ are the same. If $g_1$ and $g_2$ are encoded by a set of red  and blue edges on $\mathcal M_\beta$, respectively, they have the same probability if and only if the same multiset of parameters $\beta_i$'s are covered by the red and blue edges. In other words, the red edge set and the blue edge set are graphs with the same degree sequence. 
This degree sequence  correspondence for $\mathcal M_\beta$ seems trivial; remarkably, it holds for all linear ERGMs. 

\begin{definition}
Let $E$ be a multiset collection of edges in a hypergraph $H$.  $E$ is \emph{balanced} 
with respect to a given bicoloring of $H$ if for each vertex $v$ covered by $E$, the number of red edges containing
$v$ equals the number of blue edges containing $v$.  
 In other words, the `blue degree' of $E$ equals its `red degree'. 
\end{definition}
The  definition above appears as Definition 2.7 in \cite{PS} for the case of uniform hypergraphs, and Definition 5.7 in \cite{bouquetAlgebra} for an arbitrary hypergraph. 
 The motivation for this definition and interpretation of the binomials  as bicolored edge sets was to   study toric algebra of hypergraphs by generalizing the  algebraic literature on  graphs. The bicoloring construction idea simply generalizes Villarreal's \cite{Vill95} `closed alternating walks' on graphs, see \cite{RTT12} for a further classification and characterization of the walks. 
A straightforward argument now shows the following. 
\begin{theorem}[See Theorem 2.8  of \cite{PS} for the uniform case]
Let $\mathcal H_\mathcal M$ be the parameter hypergraph of the model $\mathcal M$. Then any balanced collection of edges $E\subset E(\mathcal H)$ constitutes a 
  move in the toric ideal associated to $\mathcal H$. In particular, the set of \emph{balanced bicolored subgraphs} connect all fibers of the model $\mathcal M$. 
\end{theorem}
Note that a balanced bicolored hypergraph is simply a set of two hypergraphs, not necessarily simple, that have the same degree sequence! 
 Examples and structure of balanced edge sets on a hypergraph, as well as extension to the non-uniform case, can be found in \cite[\S 5.1 and 5.2]{bouquetAlgebra}. 

Thus far, we have seen how all linear ERGMs that are encoded by $0/1$ matrices $M$ - `legal' incidence matrices -  give rise to parameter hypergraphs. But what about models whose probability form is more general? 
Surprisingly, by a recent result \cite[Theorem 6.2]{bouquetAlgebra}, \emph{all} toric ideals and therefore all linear ERGMs and log-linear models for discrete data are such that there exists a hypergraph that encodes their Graver bases and preserves the complexity. 
Therefore, we arrive at the following crucial fact: 
\smallskip 
\begin{center}\noindent
\emph{The parameter hypergraph encodes sufficient statistics of any linear ERGM by a hypergraph degree sequence.}  
\end{center}
\smallskip
This further reinforces the  present author's view that discrete mathematics literature on sampling graphs and hypergraphs with fixed properties should be imported to statistics and considered a crucial tool in linear ERGM model fitting in particular. 
 The following General Problem summarizes the discussion of this section: 
 \begin{problem}[A restatement of General Problem 1 with constraints]
	Fix a general linear ERGM with model hypergraph $\mathcal H$, so that the incidence matrix of $\mathcal H$  computes the sufficient statistics vector $T(g)$ for the random graph $g$. 
	Develop an efficient algorithm that samples from the set of sub-hypergraphs of $\mathcal H$ with arbitrary fixed degree sequence. 
 \end{problem}

Of course, as discussed above, we know that  `2-switches', or  
 quadratic Markov moves,  suffice to connect all realizations of hypergraph degree sequences because they correspond to the the defining ideal of  the variety of Veronese-type corresponding to the r-th  hypersimplex \cite[Chapter 14]{St}, but intermediate steps may go through non-simple hypergraphs, that is, those that have edges with multiplicities larger than $1$.  
  Further, this construction  ignores possible forbidden edges. A related problem for the restricted class of complete $k$-uniform hypergraphs can be found in the problem collection \cite[Problem 2]{DWestOnlineProblems}: For $k\geq 4$, determine whether there exists a function $f(k)$ and a set of operations, each on at most $f(k)$ vertices, that can be used to transform one $k$-realization of a $k$-graphic sequence into another.
Even if this problem is solved, it would only impact the uniform case of the hypergraph $\beta$ model, for which $\mathcal H_\mathcal M$ is complete, that is, there are no forbidden edges. For applications, statistical  sampling constraints and forbidden edges really impact the solution. For example, the famous `bad news Theorem' for Markov bases in algebraic statistics from \cite{slimTables}  shows that Markov bases for already the simplest non-trivial model on $3$ discrete random variables, called the no-three-way interaction model, are arbitrarily complicated. What this means is that there are fibers of observable data points that require moves of arbitrarily large degree if the number of states of at least two of the $3$ variables are not bounded. For completeness, let me add also the  complementing  `good news Theorem' for Markov bases famous in algebraic statistics: \cite{HoSu} show that the Markov bases for that same no-three-way interaction model  have bounded complexity if the number of states of  two of the $3$ variables is  bounded.   Incidentally, the parameter hypergraph of the no-three-way model is $3$-uniform and $3$-partite, it is not complete: there are as many edges as in $K_{p,q}$ where $p$ and $q$ are sizes of two of its parts, and it is regular in each part. So the presence of forbidden edges in $\mathcal H_\mathcal M$ compound the issue of connecting the fiber in the worst possible way. 
However, positive results can be obtained for restricted classes of problems, references to which are scattered throughout this section. 


\section{Polytopes play an important role in ERGMs} \label{sec:polytopes}

\subsection*{The parameter estimation problem}
In the estimation problem, given  an observation of the joint states of the random variables,  the goal is  to estimate the unknown probability distribution $p_{\theta_0}$ from a model $\mathcal M$  that `best explains' the data. A basic method commonly used  is that of {maximum likelihood estimation}:  a maximum likelihood estimate (MLE) of $\theta_0$ is a parameter vector  $\hat\theta\in\Theta$  that makes the given data most likely to have been observed.
It is defined as: 
\begin{equation}\label{eq:MLE}
 \widehat{\theta} = \mathrm{argmax}_{\theta \in \mathbb R^n} p_{\theta}(x).
\end{equation}
 Computing it amounts to maximizing the log-likelihood function, which maps the parameters indexing the probability distributions in a model to the likelihood of observing the given data.  
While there are several estimators that can be used, an  MLE is a very straightforward one that enjoys nice statistical properties such as consistency and asymptotic efficiency, and computing it is a seemingly simple optimization problem. 
However,  there are several obstacles.  
One if them is that the MLE of the natural parameters of the model may not exist, meaning that we are unable to make inferences about any or a subset of the parameters from the given data. 
As expected, the estimation problem has been studied in detail in the statistics literature for many  classes of models. In the last decade, it has  been shown that for discrete exponential families, the geometric structure of the model captures important information about parameter estimation including MLE existence; see \cite{RFZ:09}, \cite{DobraEtAl-IMA}, cf.\  \cite{G09}, \cite{H:03}. Luckily, in case of exponential families,\footnote{The general case is not so lucky at all:
an MLE may not exist because the likelihood function is not globally bounded; further, if an MLE exists, it may not be unique. Also, the likelihood function may be highly multimodal; a simple example is provided already by a standard example of  mixture of normal distributions. 
 In that case, standard tools such as hill-climbing algorithms (e.g.\ Newton-Rhapson) may fail to find the maximum and converge to a local optimum only, and would not be aware of it. 
Finally, MLE may be located on the boundary of the model, which is poorly understood except in  few cases in very recent literature \cite{RobevaBdry}.} 
if the MLE exists it is unique.
The following polyhedral object captures the MLE information in exponential family models. 
\begin{definition} 
 Let 
$\mathcal{T}=\{T(g)\in\mathbb{R}^d\}$ be the range of sufficient statistics $T$, where $g$ ranges over the set of all observable networks in the sample space. 
The \emph{model polytope}  associated with the family $\mathcal M$ is 
\[
\mathrm{P} = \mathrm{convhull}(\mathcal{T}) \subset \mathbb{R}^d. 
\]
We may assume $P$ to be full-dimensional. 
\end{definition}
\begin{example}\label{ex:polytopeBeta}
	The model polytope $\mathrm P$ for the $\beta$ model for  random graphs on $n$ vertices is the polytope of degree sequences of graphs on $n$ vertices. 
	\cite{RPF:11} use the explicit description of facets from \cite{MahadevPeled} to give a characterization of when the sufficient statistic vector is on the  boundary of the polytope, and study the statistical consequences.
\end{example}
The  complexity  of the MLE problem for the model $\mathcal M$ is captured by the geometry of the  polytope $\mathrm{P}$ and by the combinatorics of its face lattice as follows: 
\emph{the MLE for the observed data exists if and only if the observed sufficient statistic vector lies in the relative interior of the model polytope $\mathrm P$.} 
For instance,  \cite{RPF:11} studies MLE existence for the $\beta$ model for graphs and links to the graph theory literature on degree sequences. 
Nonetheless, partly because the link with polyhedral geometry has remained largely unexplored in both the mathematical and statistical literature, methodologies for estimation and model validation under a non-existent MLE  with proven statistical performance have yet to be developed. 
A first step in this direction is to solve the following. 

\begin{problem}[General Problem 2]\label{prob:GeneralPolytope}
	Fix a random graph model with sufficient statistics vector $T(g)$ and sample space $\mathcal G_n$. Often, $\mathcal G_n$ will be the set of simple graphs on $n$ vertices. Determine the model polytope $P=\conv\{T(g):g\in \mathcal G_n\}$.  What are the extreme points? What is the facial structure? What is the proportion of realizable lattice points on the boundary vs.\ in the  interior?  
\end{problem}
The following can be thought of a subproblem of \ref{prob:GeneralPolytope}, but due to its difficulty and relevance for developing algorithms to detect MLE non-existence, we single it out. 
\begin{problem}	[A sequel to General Problem 2]
Develop an efficient algorithm that can determine whether a given point lies on the relative interior of  $P$; i.e., obtain a facet description of the model polytope that is `efficient'. 
\end{problem}
While the general theoretical link between statistical estimation and polyhedral geometry is in place for discrete exponential families  as outlined in \cite{RFZ:09} and references therein, statistics now relies on discrete methods to provide the required tools. Let us gather here some of the relevant results. 
We emphasize again that for discrete exponential families if an MLE exists then it is unique. 
 Estimation algorithms will behave well when off the boundary of the polytope. For example, \cite{betaHypergraphs} define the hypergraph $\beta$ model and show that fixed point algorithms will converge to the MLE geometrically fast as long as the sufficient statistic is  on the relative interior of the polytope.  
\cite{RPF:11} use the polytope of graphical degree sequences to offer a complementary study to \cite{CDS11} of the estimation problem in the $\beta$ model. 
Another simple of interest is called the edge-triangle model. Its sufficient statistic vector consists of two numbers: the number of edges and the number of triangles in the graph.  Its polytope and estimation issues are addressed in \cite{RFZ:09},  the asymptotics of the MLE existence problem, polytope and its fan in \cite{AleAsymptoticQuantization}, while \cite{ZoltanSzabEvaET} study its degeneracy from the point of view of statistical mechanics, dealing with the issue of non-realizable points that exist within the model polytope in such a way that they are meant to represent data averages but are by definition non-interpretable. 

However, Problem~\ref{prob:GeneralPolytope} has not been solved for most models listed in this review. For example, the polytope of hypergraph degree sequences has been studied in the literature and some partial results are known: 
\cite{HyperDegSeqNonconvex} shows that the set of hypergraph degree sequences for uniform hypergraphs is non-convex, and thus Erd\"os-Gallai-type theorems do not hold. \cite{HyperDegSeqPolytope} also offers a very nice review of the problem, and shows that vertices of the polytope are known by Theorem 2.5. Namely, they are $r$-threshold sequences, where $r$ is the uniformity of the hypergraph. Further, \emph{some} of the facet inequalities are known. 


\subsection*{Data privacy and noisy sufficient statistics} 
Statistical analyses of network data sometimes require  projection of a noisy degree sequence  onto the model polytope, in particular its realizable points. The noisy sequence need not be a realizable degree sequence vector at all. The example below is considered  from the point of view of data privacy and confidentiality, but note that the problem is also relevant for noisy sampling, for example when there are measurement errors and the reported observed sufficient statistic does not represent a realizable sequence. 

The basic concept of data privacy and confidentiality is to minimize the risk of releasing sensitive information about, say in our case, people in a social network, while maximizing statistical utility of the data. 
Having statistical utility means that the released information can be used for statistical analyses. A straightforward example of such information that one may want to release is the observed value of the sufficient statistic. However, in the age where information about any particular data set may be gathered from different sources, releasing a sufficient statistic, e.g., the degree sequence in the $\beta$ model case, may not be `private enough'. For this reason, in the interest of preserving data privacy, noise is added to the observed sufficient statistic in order to have it releasable to the public while not jeopardizing sensitive network information.  Of course this presents various problems for statistical inference; thus the noise is added in a principled way. There are several well-known privacy-preserving mechanisms. The statistical task is then to determine how to perform reliable inference using the noisy statistic; for an example and overview of the relevant privacy literature for ERGMs see \cite{SesaVisheshPrivateERGMs}. 
One of the steps in \cite{SesaVishesBetaPrivacyAOS} that is required for computing a private parameter estimator of the released data,  with good statistical properties, is to project the noisy sufficient statistic onto the lattice points in the model polytope. 
To this end, one needs to understand the following. 
\begin{problem}[General Problem 3]\label{prob:GeneralProjectSuffStats}
	Fix a random graph model with sufficient statistics vector $T(g)$, sample space $\mathcal G_n$. Often, $\mathcal G_n$ will be the set of graphs on $n$ vertices. Let  $P=\conv\{T(g):g\in \mathcal G_n\}$ be the model polytope. 
	Develop an (efficient) algorithm to compute the projection of  a noisy sufficient statistic vector onto the realizable vectors in $P$. 
\end{problem}

Obviously, determining if a lattice point is a realizable sufficient statistic for the model $\mathcal M$ is a problem on its own. In fact realizability results of Havel-Hakimi \cite{Havel, Hakimi} type can be used to solve Problem~\ref{prob:GeneralProjectSuffStats}: 
\begin{example}
For the $\beta$ model, \cite{SesaVishesBetaPrivacyAOS, SesaVishesBetaPrivacy} use the Havel-Hakimi decomposition to compute the realizable lattice point in the polytope of degree sequences that is closest to the given noisy sequence. 
In order to achieve a more efficient implementation, they in fact use  the polytope of degree partitions \cite{AmitavaDegPartPoly}, that is, of sorted degree sequences. This polytope  can be thought of as an asymmetrized version of the degree sequence polytope. 
\end{example}
This directly leads to the analogous problems for the other models, which are singled out for convenience. 

\begin{problem}[An instance of General Problem 3]
	Generalize Havel-Hakimi to the $p_1$ model polytope. 
\end{problem} 
Remarkably, the directed part of this problem has been solved in \cite{DirectedHavelHakimiErdMiklToro}, where they also study the case of some forbidden edges. 
The authors there also prove a result useful to design an MCMC algorithm to find random realizations of prescribed directed degree sequences, which relates back to Problem~\ref{prob:p1moves} (solved in \cite{PRF10}).
However, this sampling algorithm offers a partial solution to the problem in the general case, though it should be easy to modify it to fit the statistics framework. Namely,  in the general variant of the $p_1$ model, reciprocated edges are of the form  $i\rightarrow j$ and $i\leftarrow j$ simultaneously combined into one bidirected edge $i\leftrightarrow j$. The sufficient statistics of the $p_1$ model include the number of reciprocated edges! This means that in sampling or finding random realizations, this number also matters: reciprocated edges are allowed to be `pulled apart' while sampling, but the total number of them is to remain constant. 
Recall again that forbidden edges in this case correspond to structural zeros in the $p_1$ model. 


\begin{problem}[An instance of General Problem 3]
	Generalize Havel-Hakimi to the three variants of the $\beta$ hypergraph model polytope:  uniform, layered or general hypergraph degree sequences.  
\end{problem}
The uniform instance of this problem is essentially \cite[Problem 1.1]{HyperDegSeqBehrEtAl13}, who also emphasize that 
the only known characterization of $k$-graphic sequences is due to Dewdney in 1975, but it does not  yield an efficient algorithm. Two of the variants of the $\beta$ hypergraph model work with nonuniform hypergraphs. 


\section{Concluding remarks}

There exist, of course, families of ERGMs beyond the ones discussed in this brief overview, including those based on global summary statistics not related to node degrees, for example the cores decomposition \cite{shellERGM}, or other types of models such as graphical models for networks \cite{AleKayvanGraphicalNtwks}.  
A uniform sampler for the fibers of the former model family has not been developed, while the algebra and geometry of the latter model family has not been explored. 
Clearly these are interesting problems in the statistical analysis of networks and, in particular, algebraic statistics. 
  But the main points of this chapter are:\begin{itemize}\item Several basic examples of linear ERGMs already offer many interesting open problems in discrete mathematics, instances and special cases of which are already known and interesting in their own right from the graph-theoretic point of view. \item These models also provide a theoretical foundation for exploring distributions of various network summary statistics that could serve to develop a rigorous testing and model fitting framework for networks, nowadays generally lacking despite the exploding literature on network analysis. \end{itemize}


The first type of problem discussed here is the sampling problem given a fixed set of characteristics of a network. Within the context of statistical modeling of networks, the value of sampling and therefore of exact testing  can be summarized as follows. 
Comparing to the reference set $\mathcal F_t$ 
 avoids the use of asymptotics, offers an alternative to model fit testing, and makes it unnecessary to use large-sample approximations to sampling distributions, in particular when their adequacy has {not} been determined. 
As demonstrated  on several  cases in the literatue, 
tools from algebraic statistics offer a valid and critical  approach for analyzing such models. 
However, within this realm of mathematical research, 
several   problems  remain, including problems related to scalability and applicability of the algebraic  methods. 
The second type of problem discussed here is related to the model polytope defined to be the convex hull of all sufficient statistics for the model. The polytope captures the difficulty of parameter estimation and MLE existence and is a crucial step in statistical inference. Both problems are key steps in the general scheme of testing goodness of fit of the model and model selection, detailed statistical considerations of which are beyond the scope of the present chapter.

\medskip 
To close, it is worth pointing out that algebraic statistics is  in fact a much broader field, not confined to the study of log-linear models for discrete data.  Indeed, 
the field 
 builds upon the rich and long history of the use of algebraic tools in statistics, starting with  Fisher \cite{Fisher} and  algebra for confounding \cite{PistoneWynnBiometrika}, but has expanded to encompass at least three distinct fields of mathematics: commutative and computational algebra, via solving systems of polynomial/rational equations; combinatorics of graphs, hypergraphs, and simplicial complexes; and geometry, algebraic,  convex, and polyhedral.  
The field  is roughly  two decades old, yet it has many facets,  and recent theoretical advances show that there is a significant impact on applicability and behavior of various  statistical methods, 
 including but not limited to  the following problems: parameter estimation and reliability of inference 
 for discrete exponential families  
 \cite{RFZ:09, RPF:10} and Gaussian models \cite{PiotrCaroDon}; 
 general methods for experimental design \cite{PistoneWynnEva}; 
parameter   identifiability studied in numerous references  of \cite{AlPeRhSu}, \cite{FoyDrSu} and  complexity of parameter estimation for Gaussian graphical models \cite{Mathias}, \cite{GDP}, \cite{GrSuMLthresh}; 
Bayesian model selection and singular learning theory \cite{WatanabeJMLR, WatanabeBook}, sampling from conditional or marginal distributions on contingency tables with implications to cell bounds and data privacy \cite{Sesa, SlaZhuPet}, exact tests for marginal table models \cite{Dob03}, and model fitting via exact tests for degree-based network models \cite{OHT:11, betaHypergraphs}, as well as offering new models and algorithms for networks not based on degrees \cite{shellERGM}. 
This bibliography is far from complete, but should give enough information on some recent advances in the field to the interested reader.


\section*{Acknowledgements} 
Supported in part by AFOSR Grant \#FA9550-14-1-0141. 

The author is endlessly grateful to Stephen E. Fienberg  for his guidance in the study of statistical network models, and Alessandro Rinaldo, Despina Stasi and Elizabeth Gross for continued enlightening collaborations and motivating discussions.

Many thanks to Tobias Windisch for a thorough reading  and many important corrections of the initial version of this manuscript. In addition, two anonymous referees have helped improve the text tremendously. The author is also  grateful to  Zolt\'an Toroczkai and \'Eva Czabarka for  introduction to the  relevant graph theory literature, and P\'eter L. Erd\"{o}s and Istv\'an Mikl\'os for clarifying questions and remarks.

\bibliographystyle{amsalpha}
\bibliography{AlgStat,DynMarkP1,privacy,kCoresERGM,discreteMethodsAlgStatSurvey}

\end{document}